\documentclass[journal]{IEEEtran}
\usepackage{graphicx}
\usepackage{epsf}
\usepackage{mathtools}
\usepackage{amssymb}
\usepackage{amsmath, amsfonts}
\usepackage{latexsym}
\usepackage{multirow}
\usepackage{multicol}
\usepackage[table]{xcolor}
\usepackage{color}

\usepackage{tabularx}
\usepackage{lipsum}

\usepackage{algorithm}
\usepackage{algorithmic}

\usepackage{mathrsfs}

\usepackage{fixltx2e}

\usepackage[mathscr]{euscript}

\usepackage{commath}
\usepackage{MnSymbol}

\usepackage{subcaption}

\usepackage{mathtools} 
\DeclarePairedDelimiterX\setc[2]{[}{]}{\,#1 \;\delimsize\vert\; #2\,}
\DeclarePairedDelimiterX\parth[2]{(}{)}{\,#1 \;\delimsize\vert\; #2\,}

\PassOptionsToPackage{hyphens}{url}
\usepackage[unicode=true, bookmarks=true, bookmarksnumbered=true, bookmarksopen=true, bookmarksopenlevel=1, breaklinks=false, pdfborder={0 0 0}, pdfborderstyle={}, backref=false, colorlinks=false]{hyperref}

\usepackage{theorem}
{
\theorembodyfont{\rmfamily}

}

\usepackage[nospace]{cite}

\usepackage{soul}

\usepackage{bbm}

\definecolor{orange}{RGB}{255,127,0}
\definecolor{blue}{RGB}{0,0,255}
\definecolor{red}{RGB}{255,0,0}
\definecolor{green}{RGB}{50,160,50}
\definecolor{grey}{RGB}{125,120,125}
\definecolor{purple}{RGB}{125,0,125}

\begin{document}
{
\title{
{\fontsize{16}{1}\selectfont Is Wireless Bad for Consensus in Blockchain?}
}

\author
{
Seungmo Kim, \textit{Senior Member}, \textit{IEEE}

\thanks{S. Kim is with the Department of Electrical and Computer Engineering, Georgia Southern University in Statesboro, GA, USA and can be reached at seungmokim@georgiasouthern.edu.}
}

\maketitle
\begin{abstract}
This paper examines how wireless communication affects the performance of various blockchain consensus mechanisms, focusing on their scalability and decentralization. It introduces an analytical framework for quantifying these effects, backed by extensive simulations, underscoring its broad applicability to various consensus mechanisms despite wireless communication's unreliability.
\end{abstract}


\section{Introduction}\label{sec_intro}

\subsubsection{Motivation}
In the dynamic landscape of modern connectivity, the proliferation of wireless technologies has become nothing short of permeating every facet of our daily lives and redefining the way we connect, communicate, and consume information. In fact, over 55\% of website traffic comes from mobile devices, and 92.3\% of internet users access the internet using a mobile phone \cite{mobile}. In fact, a wide variety of wireless technologies immerse our daily lives, including Wi-Fi \cite{mine_lett17}, cellular technologies such as the fifth-generation (5G) cellular \cite{mine_jsac17} and the Long-Term Evolution (LTE) \cite{mine_psun}, to vehicle-to-everything (V2X) communications \cite{mine_dsrc_arxiv20} and even wearable technologies \cite{mine_wearable_arxiv19, mine_wearable_arxiv20}.

This makes a strong case where \textit{increasingly more blockchains will be established on wirelessly connected nodes}. One should notice of the key challenge here: the reliability of wireless communications is generally lower in comparison to wired communications (e.g., ethernet), attributed to various random factors such as noise, interference, and limited bandwidth \cite{wireless}. One should also note that this lower reliability affects the performance of consensus in blockchain \cite{wireless_blockchain_mag22, wireless_blockchain_iot22}.

All open public blockchains are based on the idea that they should be able to reach consensus across a distributed network, even when there are conflicts, without putting control in one place \cite{visa}. Nonetheless, the additional dynamicity that is brought by the wireless networking has not been thoroughly discussed in the literature as of yet. Therefore, it will be a valuable academic attempt to formally analyze the performance of a blockchain system established on a wireless network.

\subsubsection{State of the Art}\label{sec_related}
PoW \cite{nakamoto08} and PoS \cite{eth_pos} remain the two most common mechanisms, especially in the cryptocurrencies' context \cite{powpos}. PoW is a form of cryptographic proof in which one party (the prover) proves to others (the verifiers) that a certain amount of a specific computational effort has been expended \cite{pow_first}. The main purpose of PoW is to deter manipulation of data by establishing large energy and hardware-control requirements to be able to do so, which inevitably leaves the it criticized by environmentalists for their energy consumption. Meanwhile, as an effort to avoid such high computational cost that PoW causes, PoS is a type of consensus mechanism for blockchains that are designed to elect validators in proportion to their quantity of holdings in the associated cryptocurrency.

Moreover, there have been proposed a variety of techniques as an effort to improve the performance of blockchains: namely, parallel structure \cite{trilemma}, off-chain \cite{offchain}, reinforcement learning \cite{mine_tiv23}, and combination of PoW and PoS \cite{trifecta}. Nonetheless, the \textit{blockchain trilemma} sets a limit on this desire: no blockchain can achieve improvements on all three fronts of scalability, security, and decentralization at once. In fact, despite the large research and experimental effort, all known approaches turn out to leave tradeoffs \cite{trilemma}.

This makes a compelling case where we urgently need a comprehensive framework evaluating the performance of blockchains with \textit{wirelessly connected} nodes with respect to the blockchain trilemma. Many of the current consensus algorithms already consider the possibilities that nodes leave, and new nodes join. However, we claim that more is needed: due to the uncertainty in wireless connections, a blockchain system established on a wireless network will draw a completely different environment than a one with ethernet-connected nodes \cite{wireless_mdpi22}.

Before our work, diverse avenues of recent research was sparked, exploring topics such as blockchain performance in the context of IoT \cite{wireless_sensors20}, communication resource consumption \cite{wireless_mag21}, efficient consensus mechanisms among wireless nodes \cite{wireless_aircon23}, connectivity-adaptive contract mechanisms \cite{wireless_iot23}, and Byzantine fault-tolerant (BFT) consensus among wireless nodes \cite{wireless_survey24}. Specifically, the existing analytical framework (e.g., universal scalability law \cite{usl_zk}) overlooked the impacts of wireless connectivity, which may cause serious imprecision these days with such a high proportion of wireless technologies in any given network.

Furthermore, most of the previous work in the literature of blockchain and distributed systems present experimental results, which can only be applied to certain practical scenarios with a set of specific parameter settings. Therefore, this paper attempts to provide a more general framework that can be applied to any other blockchains for measurement of their consensus performance.

\subsubsection{Contributions}
Addressing the aforementioned limitations of the existing literature, this papers sheds light on precisely evaluating the performance of consensus mechanisms when the nodes are wirelessly connected. The specific technical contributions are summarized as follows:
\begin{itemize}
\item It provides an analytical framework for calculation of \textit{scalability} and \textit{decentralization} of a blockchain consensus process;
\item It draws \textit{probabilitistic analysis} for formulating the impacts of wireless connections on the scalability and decentralization.
\item It presents a \textit{comparitive study} among PoW, PoS, vs PoC with respect to scalability and decentralzation.
\end{itemize}

\begin{figure}[t]
\centering
\includegraphics[width = \linewidth]{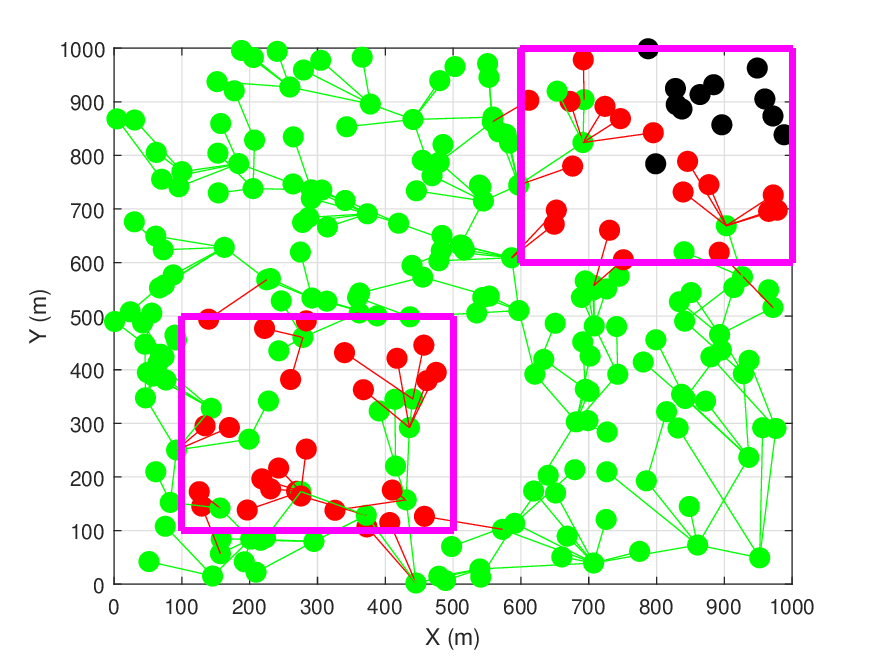}
\caption{Example scenario of gossip protocol among 400 nodes with $\mathsf{p}_{\text{fail}}=0.5$}
\label{fig_model_gossip}
\end{figure}

\section{System Model}\label{sec_model}
The key focus of this paper is the nodes being wirelessly connected. As such, the system model is supposed in such a way that the impacts of wireless connection on the blockchain performance can be clearly characterized.

The nodes are connected to each other via wireless communications. Each node holds a 100 m of communication range. This paper assumes a general ad-hoc network in which no central mediator exists; instead, the nodes are connected to their own neighbors, e.g., device-to-device communications \cite{prose}. The rationale is that a centralized network (e.g., cellular, Wi-Fi, etc.) will likely employ a permissioned blockchain (or a private blockchain), whose performance is not exactly what we are interested in measuring.

As a means to accomplish finality of data among nodes, distributed systems often adopt a \textit{gossip protocol} for propagation of information \cite{gossip}. Fig. \ref{fig_model_gossip} shows an example of the gossip protocol operated on a wireless ad-hoc network. Each node propagates the block to all the neighboring nodes within its communication range. Each connection is supposed to cause a Byzantine fault at the rate of $\mathsf{p}_{\text{fail}}$, which is shown as a red line. Such a failed connection yields a failed reception at the receiver node, which is drawn as a red dot.

A two-dimensional space $\mathbb{R}^2$ is defined as a 1 km-by-1 km square, as illustrated in Fig. \ref{fig_model_gossip}. We model spatial distribution of the stationary nodes as a Poisson point process (PPP) in $\mathbb{R}^2$ with density $\lambda$. Fig. \ref{fig_model_gossip} visualizes an example distribution of nodes with $\lambda = 400$. Notice that the $\lambda$ is the parameter that will be varied according to the type of consensus mechanism. For instance, PoS has a smaller $\lambda$ as compared to PoW since its consensus is operated only among the validators who are with more than a required amount of stake.

\begin{figure*}
\begin{minipage}[t]{0.49\linewidth}
\centering
\includegraphics[width = \linewidth]{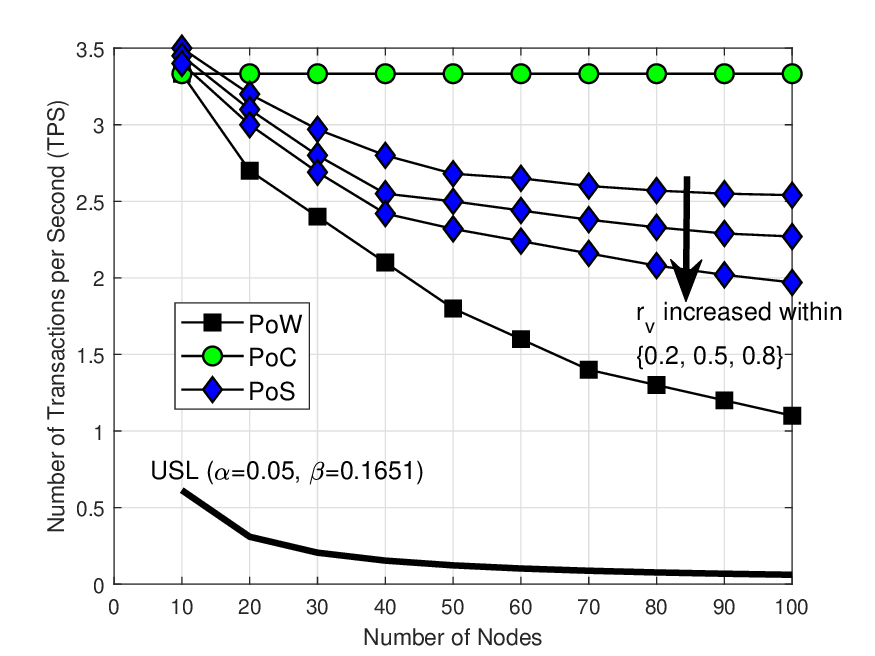}
\caption{Scalability $\textsf{R}$ versus the number of nodes ($\mathsf{p}_{\text{fail}}=5\%$, 10 transactions per block)}
\label{fig_R_all}
\end{minipage}\hfill
\begin{minipage}[t]{0.49\linewidth}
\centering
\includegraphics[width = \linewidth]{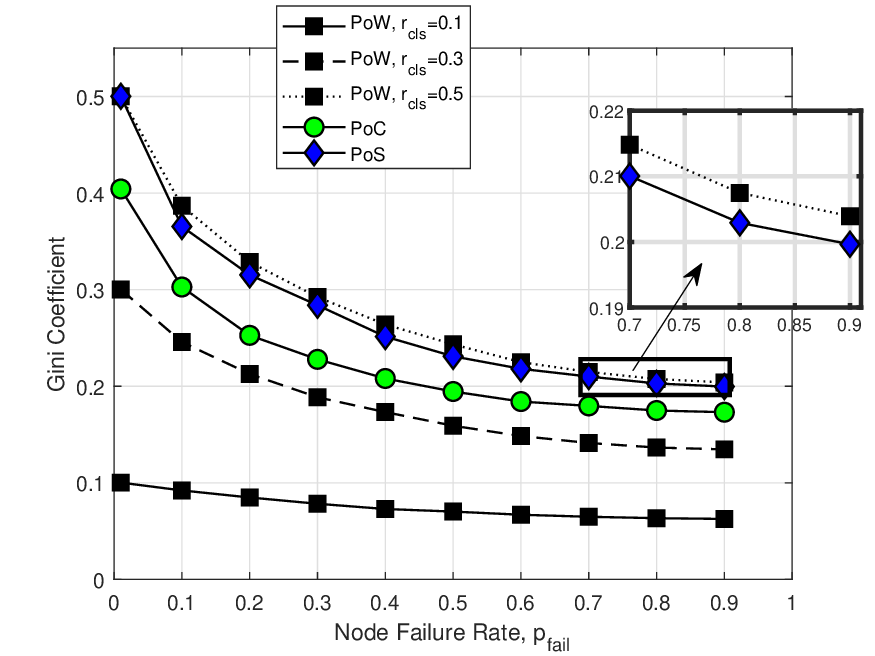}
\caption{Decentralization level $\textsf{G}$ vs. node failure rate (PoW--$\mathsf{r}_{\text{cls}}=\{0.1, 0.3, 0.5\}$; PoS--$\mathsf{r}_{\text{v}}=0.2$; PoC--$\mathsf{r}_{\text{sfl}}=0.9$, $\Delta_{\text{sfl}}=0$)}
\label{fig_gini_all}
\end{minipage}
\end{figure*}

\section{Consensus among Wireless Nodes}
Node failures occur in a ``\textit{clustered}'' manner; the magenta-colored boxes indicate the areas where the failures transpire. It is assumed that the node failures take place within the clusters only; the example presented in Fig. \ref{fig_model_gossip} assumes the probability of each node failure to be $\mathsf{p}_{\text{fail}}=0.5$. We will vary this $\mathsf{p}_{\text{fail}}$ as a method to evaluate the performance of the consensus mechanisms--viz., PoW, PoS, and PoC.

The key discussion of this paper is that the wireless connection can increase the odds of Byzantine faults. Specifically, we identify \textit{``clustered occurrence'' of faults attributed to interference among each other} \cite{haenggi_survey} as the key characteristic that the wireless connection adds to Byzantine faults, which yields that the nodes located in certain areas face a higher probability of disconnection. We attribute such clustered disconnections to \textit{interference among nodes}. In fact, an ad-hoc wireless network is more susceptible to interference \cite{adhoc_interference}, due to its decentralized network structure where no central mediator exists to coordinate scheduling among nodes.

Specifically, we define the consensus mechanisms that we analyze in this paper--i.e., PoW, PoS, and PoC--and identify their key characteristics that may be affected by the wireless connections. Moreover, it is noteworthy that we pay attention to PoC and compare it to PoW and PoS, the two most predominating consensus mechanisms. The rationale is that PoC adopts a significantly different principle from that of PoW and PoS, which thus distinguishes it from the vast majority of recent consensus mechanisms built on PoW and PoS.

\section{Analysis on Consensus Performance}\label{sec_analysis}
Now, we present the theoretical analysis framework that this paper proposes, which features two metrics measuring the performance of a consensus mechanism.

\subsubsection{Scalability}\label{sec_analysis_scalability}
The literature of distributed systems often rely on the \textit{universal scalability law} \cite{usl, scalability_survey} as the key scalability measure. Nonetheless, this quantity is usually obtained \textit{empirically}, which makes it almost impossible to analytically quantify $\alpha$ and $\beta$. Moreover, we could find an extensive empirical analysis of the parameters for Zookeeper \cite{usl_zk}, one of the popular distributed systems \cite{zk}. In fact, Zookeeper has appeared as on many of the blockchain systems \cite{zk_iot, zk_hyperledger}. This makes a case that we adopt the result of $\textsf{S}$ found for the Zookeeper as a \textit{scalability benchmark} for this paper's result.

Motivated from the challenge, we define a generalized metric for the scalability in this paper, which is given by
\begin{align}\label{eq_R}
\mathsf{R} = \displaystyle \frac{\text{\# transactions}}{\text{\# seconds}} = \frac{\mathsf{n}_{\text{tx}}}{\mathsf{T}_{c}}
\end{align}
where $\mathsf{T}_{c}$ denotes the length of time for a single consensus.

\subsubsection{Decentralization}\label{sec_analysis_decentralization}
We propose to adopt \textit{Gini coefficient} (denoted by $\textsf{G}$) to represent the level of decentralization. To elaborate, we modify $\textsf{G}$ in such a way that the ``inequality'' indicates the level of participations being concentrated to a fewer number of nodes, which is understood as ``centralized.''

Building on this understanding, one can take a more formal way to write this decentralization analysis. We define the Gini coefficient as half of the relative mean absolute difference, which is equivalent to the definition based on the Lorenz curve \cite{gini_econ}. The mean absolute difference is the average absolute difference of all pairs of items of the population, and the relative mean absolute difference is the mean absolute difference divided by the average, $\bar{x}$, to normalize for scale. If $x_{i}$ is the number of node  $i$'s parcipations in consensus, and there are $n$ nodes, then $\textsf{G}$ is given by
\begin{align}\label{eq_G}
\textsf{G} = \frac{\displaystyle \sum_{i=1}^{n}\sum_{j=1}^{n}\left|x_{i} - x_{j}\right|}{\displaystyle 2\sum_{i=1}^{n}\sum_{j=1}^{n}x_{j}} = \frac{\displaystyle \sum_{i=1}^{n}\sum_{j=1}^{n}\left|x_{i} - x_{j}\right|}{2n^2 \bar{x}}.
\end{align}

\section{Numerical Results}\label{sec_results}
Fig. \ref{fig_R_all} compares the scalability, $\mathsf{R}$, among the three consensus mechanisms--i.e., PoW, PoS, and PoC. One can easily find that PoC is the most scalable among the three since it is designed to select a fixed number of witnesses, which also explains the ``flatness'' of its scalability versus the number of nodes. Due to such selectivity, PoC should pose a lower level of decentralization, as shall be presented in Fig. \ref{fig_gini_all}.

The scalability of PoS depends on $\mathsf{r}_{\text{v}}$, indicating the \textit{rate of validators}. We set the parameter as a Poisson random variable with a justification that there must be a ``certain amount'' of tokens that the largest number of participants hold.

Fig. \ref{fig_gini_all} indicates that a higher wireless disconnection rate increases the decentralization level. One should also notice that the three consensus mechanisms pose the order of PoW $>$ PoC $>$ PoS in terms of the decentralization level. For PoS, the rate of validators among all the nodes is set to $\mathsf{r}_{\text{v}}=0.2$. For PoC, the rate of node replacement in each round of consensus is set to $\mathsf{r}_{\text{sfl}}=0.9$, and the number of rounds between a replacement is $\Delta_{\text{sfl}}=0$. To wit, $\mathsf{r}_{\text{sfl}}$ gives \textit{how many of the current witnesses} will be replaced with other ones; and $\Delta_{\text{sfl}}$ means \textit{how frequently} the replacement occurs.

The most striking result is that wireless connection helps \textit{improve} the decentralization in blockchain! The reason for this result is straightforward. The intermittent wireless connection hinders such participation. Therefore, the level of ``inequality'' is decreased among nodes participating in consensuses.

The rate of cluster (denoted by $\mathsf{r}_{\text{cls}}$) was varied with PoW only. The rationale is that PoW is particularly susceptible to such clusteredness of node failure transpiration. In contrast, PoS and PoC will be less impacted since they already adopt ``selective'' participation of nodes in a consensus; to wit, if failures occur at non-participating nodes, they have almost no impact on consensus.

\section{Future Work}
We will extend the findings of this paper to analyzing how mobility \cite{mobility} affects the consensus in blockchain. Moreover, we will investigate impacts of off-chain PoC migration (as part of the recent migration to Solana \cite{poc_solana}) on this paper's result.


\end{document}